\documentclass[journal]{IEEEtran}

\usepackage{amssymb}
\usepackage{amsmath}
\usepackage{graphicx}
\usepackage{color}	

\newcommand{\dv}{d_\mathrm{v}}
\newcommand{\dc}{d_\mathrm{c}}
\newcommand{\GF}{\mathrm{GF}}

\begin{document}

%
\title{Near Capacity Approaching for Large MIMO Systems by Non-Binary LDPC Codes with MMSE Detection}
%
%

%
%
%

\author{Puripong~Suthisopapan, Kenta~Kasai, Anupap Meesomboon and~Virasit Imtawil
}


\maketitle

\thispagestyle{empty}
\pagestyle{empty}

\begin{abstract}
In this paper, 
we have investigated the application of non-binary LDPC codes to 
spatial multiplexing MIMO systems with a large number of low power antennas. 
We demonstrate that such large MIMO systems 
incorporating with low-complexity MMSE detector and non-binary LDPC codes 
can achieve low probability of bit error at near MIMO capacity.
The new proposed non-binary LDPC coded system also 
performs better than other coded large MIMO systems known in the present literature.
For instance, non-binary LDPC coded BPSK-MIMO system with 600 transmit/receive antennas 
performs within 3.4 dB from the capacity 
while the best known turbo coded system operates about 9.4 dB away from the capacity.
Based on the simulation results provided in this paper,
the proposed non-binary LDPC coded large MIMO system is capable of supporting 
ultra high spectral efficiency at low bit error rate.
\end{abstract}


\section{Introduction}

In order to support very high data rate on wireless communication channels, 
it is known that the physical limitation is bandwidth
since, at the present moment, it is very scarce and extremely expensive. 
One of the possible solutions to meet this constraint is 
the multiple-input multiple-output (MIMO) system \cite{mimo1,mimo2}.
The MIMO system is a transmission system 
that uses multiple antennas at both sides of the communication ends.
The MIMO system which utilizes spatial multiplexing technique (known as BLAST system) \cite{sm_mimo1,sm_mimo2}
has attracted a great deal of attention over the past two decades
since it provides the significant increase in spectral efficiency
without additional bandwidth and transmit power in a rich scattering environment.
The transmitter spatially multiplexes data streams 
and then simultaneously transmits these multiplexed data via different transmit antennas.
At the receiver side, some specific techniques will be employed to demultiplex the received data.

In this paper, we focus our attention to the spatial multiplexing MIMO system 
that employs tens to hundreds transmit/receive antennas 
and refer to such a system as \textit{large} MIMO system.
The prominent advantage of large MIMO system is the improving in capacity
which is proportional to the minimum of the number of transmit and receive antennas \cite{mimo2}.
As pointed out in the literature \cite{coded_large_mimo1,coded_large_mimo3},
placing a large number of antennas is more amenable
because the transmitted RF energy can be more sharply focused in space.
Therefore, we believe that a large MIMO system would be practical in the near future 
to support the increasing demand for high data rates in wireless communications.

Concatenating the MIMO system with channel codes is a methodology 
to increase reliability and performance of the MIMO system.
There has been a tremendous effort to develop the coded MIMO systems, 
which include code-design, the invention of soft-output detectors, 
joint detection and decoding techniques and so forth, 
with the ultimate goal to approach MIMO capacity (see e.g. \cite{cap_approach1,cap_approach2,cap_approach3}).
We note that almost works have studied the coded MIMO system 
with less number of transmit/receive antennas (e.g. 2 to 8 transmit/receive antennas).
Therefore, lacking in the literature is a performance study of coded large MIMO system.

To the best of our knowledge,
only large MIMO systems concatenated with turbo code have been studied in \cite{coded_large_mimo1,coded_large_mimo3,coded_large_mimo2}.
Although the turbo coded large MIMO system mentioned above 
is very attractive in term of computational complexity
but there still exists a significant performance gap to the MIMO capacity.
For example, the gap of turbo coded MIMO system with 200 transmit/receive antennas
from MIMO capacity is more than $7$ dB.
In this paper, we aim to reduce the remaining gap
by considering non-binary low-density parity-check (NBLDPC) codes \cite{nb_ldpc}.

Thank to the superior performance for short and moderate codeword lengths,
NBLDPC codes have recently received an upsurge of research interest from wireless community 
including the application to MIMO channels \cite{nb_mimo1,nb_mimo2,nb_mimo3,nb_mimo4}. 
For the MIMO system with 2 transmit/receive antennas, 
it has been reported in \cite{nb_mimo3} that
a regular NBLDPC code defined over $\GF(2^8)$ outperforms both
optimized irregular binary LDPC code 
and binary LDPC code defined in the latest IEEE standard
(the binary LDPC code is specified by parity-check matrix defined over $\GF(2)$).
However, the NBLDPC coded system in \cite{nb_mimo3} deploys the optimal maximum a posterior (MAP) detector 
to initialize the soft-input for NBLDPC decoder.
This is impossible for the large MIMO system
since the complexity of MAP detection grows exponentially with the number of transmit antennas
and the size of modulation constellation.
We also note that the performance of NBLDPC codes for MIMO systems with suboptimal detector has not been reported.

In this study, 
we have investigated the application of NBLDPC codes to large MIMO systems
which utilize the low-complexity minimum mean square error (MMSE) detection as MIMO detector. 
Our contributions in this paper can be summarized as follows :

1) The non-binary LDPC coded large MIMO system with MMSE detector is proposed in this paper.
We also present how the soft-output MMSE detector can straightforwardly work with NBLDPC decoder.

2) We provide the simulation results which can be used as the benchmark for coded large MIMO systems.
We have shown that the non-binary LDPC coded system performs best among other coded schemes at SNR close to MIMO capacity.
when a large number of antennas are employed.

The rest of this paper is organized as follows. 
We first introduce the NBLDPC codes in Section II. 
In Section III, the system model used for all simulations is described. 
In Section IV, we explain the soft-output MMSE detector for NBLDPC decoder.
In Section V, we present the decoding performance of NBLDPC coded large MIMO systems.
This paper is closed with conclusions.

\section{NBLDPC Codes}
An NBLDPC code $C$ over Galois field $\GF(2^m)$ is defined by 
the null-space of a sparse $P \times N$ parity-check matrix $\mathbf{A}=\lbrace a_{ij} \rbrace$ 
defined over $\GF(2^m)$, for $i=1,\ldots,P$ and $j=1,\ldots,N$
\begin{align*}
C = \{ \mathbf{x} \in \GF(2^m)^{N} \mid \mathbf{A}\mathbf{x}^\mathsf{T} = \mathbf{0} \in \GF(2^m)^{P}\},
\end{align*}
where $m>1$ and $\mathbf{x} = (x_1,\ldots,x_N)$ is a codeword.
The $i$-th parity-check equation can be written as
\begin{align*}
a_{i1}x_1 + a_{i2}x_2 + \cdots + a_{iN}x_{N} = 0,
\end{align*}
where $a_{i1},\ldots,a_{iN} \in \GF(2^m)$ are the entries of $i$-th row of $\mathbf{A}$.
The parameter $N$ is the codeword length in symbol.
Assuming that $\mathbf{A}$ is of full rank, 
the number of information symbols is $K=N-P$ and the code rate is $R = K/N$.

We note that a non-binary symbol which belongs to $\GF(2^m)$ 
can be represented by the binary sequence of length $m$ bits.
For each $m$, we fix a $\GF(2^m)$ with a primitive element $\alpha$ and its primitive polynomial $\pi$. 
Once a primitive element $\alpha$ of $\GF(2^m)$ is fixed, 
each non-binary symbol is given by an $m$-bits representation \cite[p.~110]{macwilliams77}.
For example, with a primitive element $\alpha\in\GF(2^3)$ such that $\pi(\alpha)=\alpha^3+\alpha+1=0$, each symbol is represented as
$0=(0,0,0)$, $1=(1,0,0)$, $\alpha=(0,1,0)$, $\alpha^2=(0,0,1)$,
$\alpha^3=(1,1,0)$, $\alpha^4=(0,1,1)$, $\alpha^5=(1,1,1)$ and $\alpha^6=(1,0,1)$.
Let $L(x)$ be the binary representation of $x \in \GF(2^m)$.
For the above example, we can write $L(x=\alpha^3) = (1,1,0)$.
Thus, each coded symbol $x_i \in \GF(2^m), \forall i \in \lbrace1,\ldots,N\rbrace$ 
of a non-binary codeword represents $m$ bits.
We also denote $n = mN$ and $k = mK$ as the codeword length and information length in bit, respectively.

An NBLDPC code is $(\dv,\dc)$-regular if 
the parity-check matrix of the code has constant column weight $\dv$ and row weight $\dc$.
The parity-check matrix $\mathbf{A}$ can be represented by 
a Tanner graph with variable and check nodes \cite[p.~75]{mct_book}.
The belief propagation (BP) algorithm  for NBLDPC decoder \cite{nb_ldpc} exchanges the probability vector
of length $2^m$ between variable nodes and check nodes of the Tanner graph  
at each iteration round $\ell$.

In this paper, only is $(\dv=2,\dc)$-regular NBLDPC code defined over $\GF(2^8)$ considered due to the following reasons :

1) The process to optimize parity-check matrix $\mathbf{A}$ is not required 
since it is empirically known as the best performing code especially for short code length.
Moreover, the NBLDPC code with $\dv=2$ can be encoded in linear time \cite{cons1}.

2) The high decoding complexity of NBLDPC decoder can be compensated
since the $\mathbf{A}$ is very sparse \cite{cons2}.

3) As stated in the introduction, 
NBLDPC code defined over $\GF(2^8)$ seems to offer 
the best performance for MIMO system with MAP detection \cite{nb_mimo3}.
We therefore expect the excellent performance of NBLDPC codes when applying to large MIMO systems.

4) It is empirically shown that the application of this code 
to higher order modulation is outstanding 
\cite[p.~32]{cons3}.

It is not overly exaggerated to state that we intend to apply 
the simple, low-complexity, high-performance channel code to large MIMO systems.

\section{System Model}
We adopt the conventional notation 
to denote the MIMO system with $N_t$ transmit antennas and $N_r$ receive antennas
as $N_t \times N_r$ MIMO system.
Let $\mathbb{A}^{M}$ be the complex modulation constellation
of size $M=2^p$ where $p$ represents bit(s) per modulated symbol.
In this study, each antenna uses the same modulation scheme and the mapping is a Gray-labelled constellation.

\begin{figure}[htb]
\centering
\includegraphics[scale=0.675]{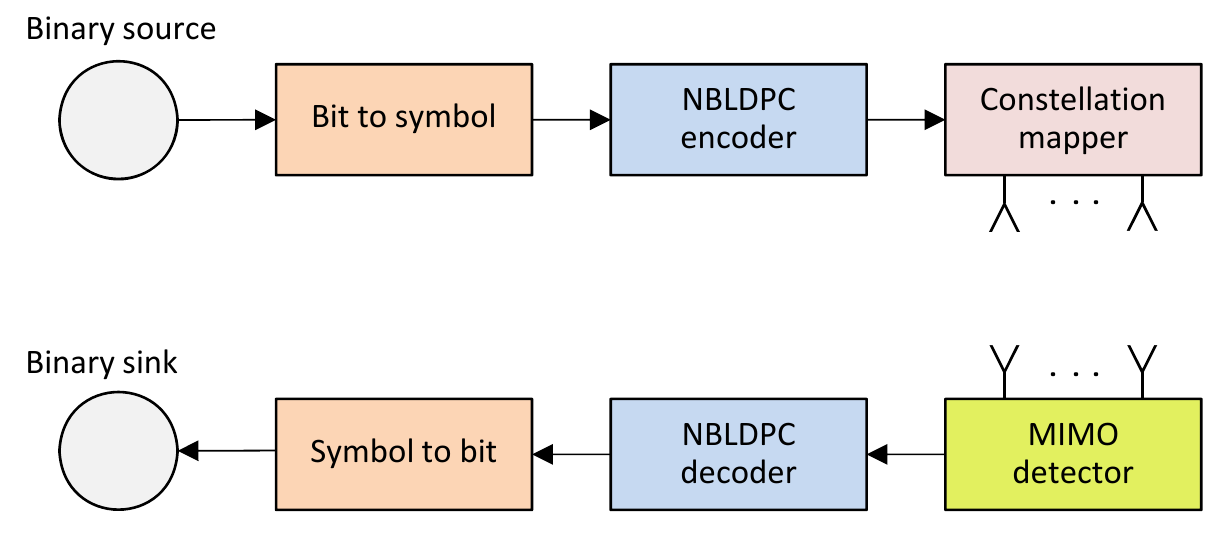}
\caption{NBLDPC coded MIMO system. Every 8 bits is mapped to a symbol in $\GF(2^8)$.}
\label{NBLDPC_coded_MIMO}
\end{figure}

Figure \ref{NBLDPC_coded_MIMO} shows the spatial multiplexing MIMO system 
concatenated with an NBLDPC code of rate $R$ defined over $\GF(2^8)$.

At the transmitter side, a bit to symbol mapper maps a group of 8 information bits 
to a non-binary symbol in $\GF(2^8)$.
The stream of $K$ non-binary symbols is encoded into a codeword of length $N$ symbols through an NBLDPC encoder.
Each coded symbol in $\GF(2^8)$ is then mapped to a group of $q=$ 8$/p$ modulated symbols by a constellation mapper.
At each time instant (each channel use), 
the transmitter simultaneously sends $N_t = K_tq$ modulated symbols in parallel 
through $N_t$ transmit antennas where $K_t$ is a number of coded symbols per each transmission.
Let $\mathbf{s} = [s_1,s_2,\ldots,s_{N_t}]^{T}\in\mathbb{C}^{N_{t}}$ be the transmit signal vector.
Each entry $s_i , \forall i \in \lbrace1,\ldots,N_t\rbrace $ taken from $\mathbb{A}^M$ must satisfy the component-wise energy constraint $\mathrm{E}[\|s_i\|^2]= E_s/N_{t}$ 
where $E_s$ is the total transmitted power and  $\mathrm{E}[\cdot]$ denotes the expectation.
With this energy constraint, a large number of transmit antennas 
imply low power consumption per each transmit antenna.

Consider a $16 \times 16$ MIMO system with QPSK modulation $(M=4 \text{ and } p=2)$ as an example.
After encoding, each coded symbol in $\GF(2^8)$ is mapped to $q=4$ modulated symbols.
At each time instant, the transmitter collects 16 modulated symbols mapping from $K_t=4$ coded symbols.
These 16 modulated symbols are multiplexed and sent through 16 transmit antennas. 

Considering a baseband discrete time model for a flat fading MIMO channel,
the received vector $\mathbf{y} = [y_1,y_2,\ldots,y_{N_r}]^{T}\in\mathbb{C}^{N_{r}}$ 
of the spatial multiplexing $N_t \times N_r$ MIMO system is given by \cite{coded_large_mimo1}
\begin{equation}
\label{model_eq}
\mathbf{y} = \mathbf{H}\mathbf{s} + \mathbf{n}.
\end{equation}
The matrix 
$\mathbf{H} = [\mathbf{H}_1 \mathbf{H}_2 \ldots \mathbf{H}_{N_t}] \in\mathbb{C}^{N_{r} \times N_{t}} $ 
denotes the channel fading matrix 
whose entry $h_{kj}$ is assumed to be complex Gaussian random variable 
with zero mean and unit variance $\mathrm{E}[\|h_{kj}\|^2]=1$.
The vector $\mathbf{n} = [n_1,n_2,\ldots,n_{N_r}]^{T}\in\mathbb{C}^{N_{r}}$ is a noise vector 
whose entry is a complex white Gaussian noise with zero mean and variance $\sigma^{2}_{n}$ per real component.
The MIMO detector performs detection and produces the prior probabilities (soft-output) for NBLDPC decoder.
After all $N$ variable nodes are initialized, the NBLDPC decoder performs decoding process
and provides the estimated non-bianry symbols (hard output).
These estimated symbols are finally demapped to a sequence of estimated information bits.
In this paper, the channel matrix $\mathbf{H}$ is assumed to be known at the receiver 
and we only focus on the square channel matrix, i.e. $N_t = N_r$.

Since each entry of $\mathbf{H}$ has unit variance, 
the average signal energy per receive antenna is $E_s$.
We follow the convention that $N_0/2 = \sigma^{2}_{n}$ to define the signal to noise ratio.
In this setting, the average signal to noise ratio (SNR) per receive antenna, 
denoted by $\gamma$, is given by \cite{cap_approach1}
\begin{equation}
\label{SNR_eq}
\gamma = \frac{E_s}{N_0} = \frac{E_s}{2\sigma^{2}_{n}}.
\end{equation}

The spectral efficiency (transmitted information rate) of coded MIMO system with spatial multiplexing technique 
is $pRN_t$ \cite{cap_approach1}.
With perfect $\mathbf{H}$ at the receiver side, ergodic MIMO capacity is given by \cite{mimo2}
\begin{equation}
\label{C_eq}
C = E\left[ \log_2\det\left(\mathbf{I}_{N_r}+ \left(\gamma/N_t\right)\mathbf{H}\mathbf{H}^\mathsf{H}\right) \right],
\end{equation}
where the superscript $\mathsf{H}$ denotes the Hermitian transpose operator, $\det$ denotes the determinant 
and $\mathbf{I}_{N_r}$ is the identity matrix of size $N_r \times N_r$.
The MIMO capacity defined in (\ref{C_eq}) can be evaluated via the Monte-Carlo simulation.
Both of spectral efficiency and capacity are measured in bits/sec/Hz (bps/Hz). 



%

\section{Soft-Output MMSE Detector \\for NBLDPC Decoder}
One of the major problems 
that prohibits large MIMO systems from practical implementation
is the high computational complexity of MIMO detector.
Even with modern and high-speed circuits, 
the optimal MAP or ML (maximum likelihood) detector is still infeasible.
Therefore, many works have recently focused on inventing the low-complexity detectors 
to enable the use of large scale multiple antennas \cite{large_mimo_detector}. 

In this paper, we consider the MMSE detector which is a famous class of linear detection
since it has low computational complexity.
Based on the complexity analysis provided in \cite{coded_large_mimo1},
the MMSE detector has an average per-bit complexity of $O(N_tN_r)$ 
which is lower than the well-known ZF-SIC (zero-forcing with successive interference cancellation) detector 
whose average per-bit complexity is $O(N_t^2N_r)$. 
Several near-optimal MIMO detection algorithms 
such as sphere detection or lattice reduction-aided detection
also exhibit high computational complexity which is much greater than $O(N_tN_r)$ \cite{coded_large_mimo1}.

Let us introduce the MMSE detector and the soft-output generation for NBLDPC decoder.
Following \cite{soft_mmse}, 
a detection estimate $\hat{s}_k$ of the transmitted symbol on $k$-th antenna
by multiplying $y_k$ with the MMSE weight matrix $\mathbf{W}_k$
\begin{equation}
\label{detect_eq}
\hat{s}_k = \mathbf{W}_{k}^\mathsf{H} y_k,
\end{equation}
where the MMSE weight matrix $\mathbf{W}_k$ is of the form
\begin{equation}
\label{MMSE_eq}
\mathbf{W}_{k} = \left( \frac{N_0}{E_s/N_t} \mathbf{I}_{N_r} + \mathbf{H}\mathbf{H}^\mathsf{H}  \right)^{-1} \mathbf{H}_k.
\end{equation}
This MMSE weight matrix $\mathbf{W}_k$ is chosen so as to minimize the mean square error 
between the transmitted symbol $s_k$ and $\hat{s}_k$.
It is important to note that direct computation of the matrix inverse in $(\ref{MMSE_eq})$ 
can be avoided by using Sherman-Morrison-Woodbury formula \cite[p.~50]{SMW_formula}

The estimation $\hat{s}_k$ can be approximated as 
the output of an equivalent additive white Gaussian noise (AWGN) channel \cite{Eq_AWGN}
\begin{equation}
\label{Eq_AWGN_eq}
\hat{s}_k = \mu_k s_k + z_k,
\end{equation}
where $ \mu_k = \mathbf{W}_{k}^\mathsf{H} \mathbf{H}_k $ 
and $z_k$ is a zero-mean complex Gaussian variable 
with variance $\epsilon^{2}_{k} = \frac{E_s}{N_t} ( \mu_k - \mu_{k}^{2} )$.
Based on this approximation, 
the probability of $\hat{s}_k$ conditioned on $s \in \mathbb{A}^{M}$ is as follows
\begin{equation}
\label{Prob_eq}
\mathrm{Pr}\left( \hat{s}_k \mid s \right) \simeq  \frac{\kappa}{\pi\epsilon^{2}_{k}} \text{exp} \left( -\frac{1}{\epsilon^{2}_{k}} \parallel \hat{s}_k - \mu_k s \parallel^2 \right),
\end{equation}
where $\parallel\cdot\parallel^2$ denotes the squared Euclidean norm and $\kappa$ is the normalized constant such that 
$\sum_{s \in \mathbb{A}^M } \mathrm{Pr}\left( \hat{s}_k \mid s \right) = 1 $.

Let $p^{(0)}_{v}(x)$ denote the input of belief propagation (BP) algorithm which is used as NBLDPC decoder
where $v = 1,2,\ldots,N$ and the index in superscript represents the iteration round of BP algorithm.
The $p^{(0)}_{v}(x)$ is the probability that $v$-th coded symbol is likely to be $x \in \GF(2^8)$.
We assume that $k$-th transmit antenna to $(k+q-1)$-th transmit antenna 
are used to send the $v$-th coded symbol (any coded symbol is mapped to $q$ modulated symbols).
We mention that the modulated or constellation symbol $s^{k} \in \mathbb{A}^{M}$ 
can be demapped to $p$ bits according to the label of constellation.
For each $x \in \GF(2^8)$, we need to collect $q$ modulated symbols to represent $qp = 8 $ bits.
The generation of soft-output from MMSE detector for
$v$-th variable node of BP algorithm is as follows.
\begin{equation}
\label{input_eq}
p^{(0)}_{v}(x) = \prod^{q-1}_{i=0}\mathrm{Pr}\left( \hat{s}_{ k+i } \mid s^{ k+i } \right),
\end{equation}
where $\hat{s}_{k+i}$ is the estimation on the $(k+i)$-th receive antenna,
$s^{k+i} \in \mathbb{A}^{M}$
and the eight bits ordered sequence of $( s^{k}, s^{k+1}, \cdots s^{k+q-1} )$ 
must be equal to the binary representation $L(x)$.
According to (\ref{input_eq}), 
$2^8(q-1)$ real multiplications for each coded symbols are needed 
to calculate $p^{(0)}_{v}(x)$ for all $x \in \GF(2^8)$.
We note that the computational complexity of generating soft-output for NBLDPC decoder is rather low
when comparing to the computation of MMSE matrix. 

For the sake of completeness, we now introduce the BP algorithm for NBLDPC decoder.
The BP algorithm mainly consists of 4 parts which can be described as follows (in this paper we use $m=8$).

\noindent\textbf{Initialization} : We set the iteration round $\ell = 0$ 
and define the maximum iteration $\ell_{\textrm{max}}$.
For each variable node $v$, $p_{v}^{(0)}(x)$ is computed from (\ref{input_eq}).
Each variable node sends the initial message $p_{vc}^{(\ell=0)} = p_{v}^{(0)} \in \mathbb{R}^{2^m}$
to each adjacent check node $c$ where $c=1,\ldots,P$.

\noindent\textbf{Check to Variable} : For each check node $c$, 
let $\partial_c$ be the set of adjacent variable nodes of $c$.
The check node $c$ sends the following message $p_{cv}^{(\ell)} \in \mathbb{R}^{2^m} $ to each adjacent variable node $v \in 
\partial_c$
\begin{align*}
&  \tilde{p}_{vc}^{(\ell)}(x) = {p}_{vc}^{(\ell)}(a_{vc}^{-1}x) \text{ for }x \in \GF(2^m),\\
& \tilde{p}_{cv}^{(\ell+1)} = \otimes_{v' \in \partial_c \setminus \{v\}  } \tilde{p}_{v'c}^{(\ell)},\\
& p_{cv}^{(\ell+1)}(x) = \tilde{p}_{cv}^{(\ell+1)}(a_{cv}x) \text{ for }x \in \GF(2^m),
\end{align*}
where $p_1 \otimes p_2 \in \mathbb{R}^{2^m} $ 
is the convolution of $p_1 \in \mathbb{R}^{2^m}$ and $p_2 \in \mathbb{R}^{2^m}$
which can be expressed as 
\begin{align*}
 ( p_1 \otimes p_2 ) (x)  = \sum\limits_{\scriptstyle y,z \in \GF(2^m ) \hfill \atop \scriptstyle ~~~x = y + z \hfill} {p_1(y)p_2(z)} \text{ for }x \in \GF(2^m).
\end{align*}
The convolution appeared above can be efficiently calculated via FFT and IFFT \cite{FFT_BP}.
Increment the iteration round as $\ell:=\ell+1$.

\noindent\textbf{Variable to Check} : For each variable node $v = 1,2,\ldots,N$, let $\partial_v$ be the set of adjacent check nodes of $v$.
The message $p_{vc}^{(\ell)} \in \GF(2^m)$ sent from $v$ to $c$ is computed by
\begin{center}
$p_{vc}^{(\ell)}(x) = \xi p_{v}^{(0)}(x) \prod_{c' \in \partial_v\setminus\{c\}}  p_{c'v}^{(\ell)}(x)$ for $x \in \GF(2^m)$,
\end{center}
where $\xi$ is the normalized constant so that 
$\sum_{x \in \GF(2^m) } p_{vc}^{(\ell)}(x) = 1$.

\noindent\textbf{Tentative Decision} : The tentative decision of the $v$-th symbol is given by 
\begin{align*}
 \hat{x}_{v}^{(\ell)} = \mathop {\arg \max }\limits_{x \in \GF(2^m )} p_v^{(0)}(x)\prod_{c \in \partial_v}  p_{cv}^{(\ell)}(x).
\end{align*}
The algorithm stops when the maximum iteration $\ell_{\textrm{max}}$ is reached 
or $\mathbf{A}\hat{\mathbf{x}} = \mathbf{0} \in \GF(2^m)^P$. 
Otherwise repeat the latter 3 decoding steps.

%
%
%
%
%
%

\section{Simulation Results}

In this section, the performance of NBLDPC coded MIMO system with MMSE detector 
(NBLDPC coded system for simplicity) is presented.
The maximum iteration of NBLDPC decoder $\ell_{\textrm{max}}$ is set to 100 for all simulation results.

\begin{figure}[htb]
\centering
\includegraphics[scale=0.535]{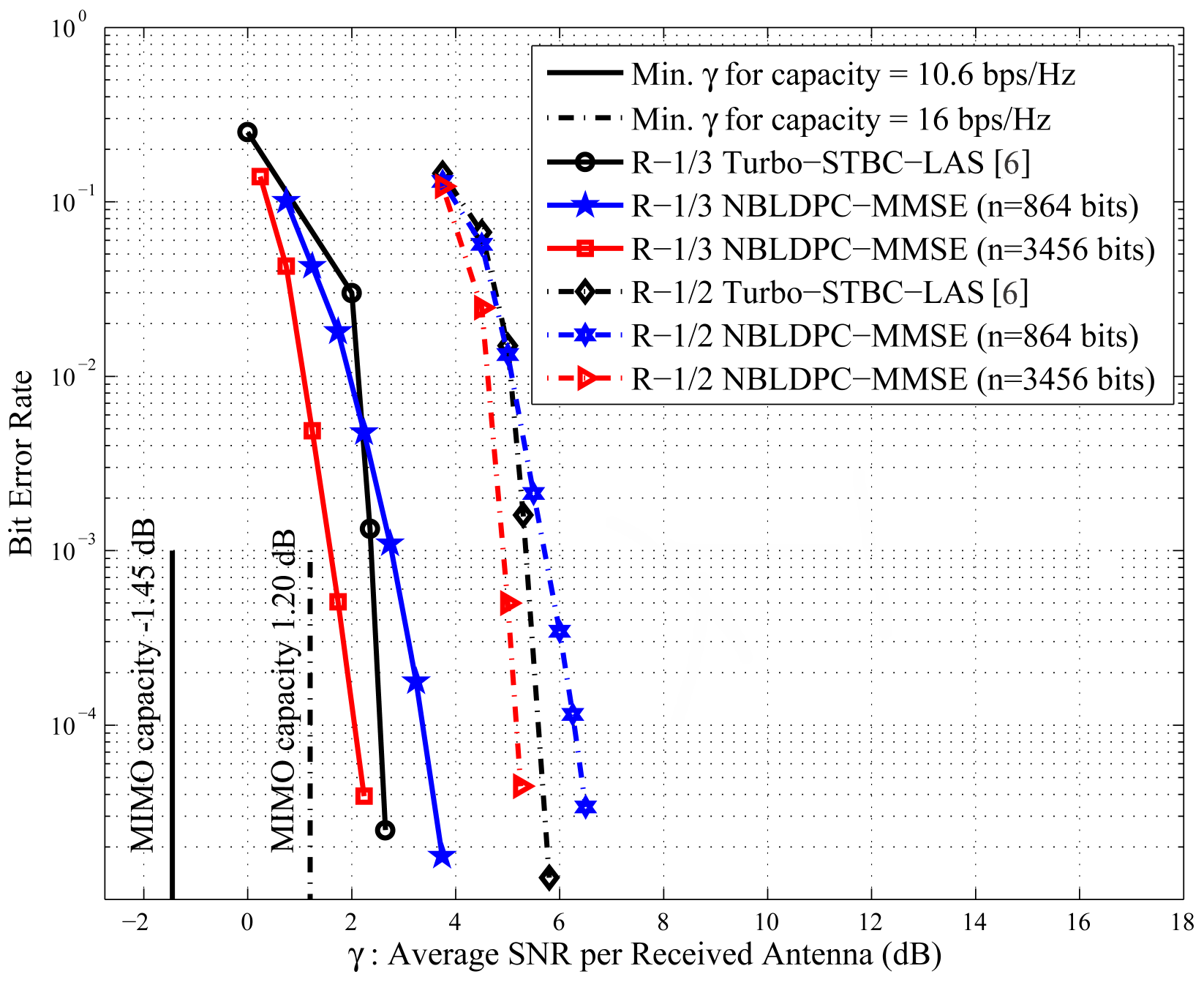}
\caption{Bit error rate curve of coded $16 \times 16$ MIMO systems with QPSK modulation.
The spectral efficiencies of 10.6 and 16 bps/Hz 
are obtained from MIMO system with channel codes of $R=1/3$ and $R=1/2$, respectively.}
\label{ber16x16}
\end{figure}

In Fig. \ref{ber16x16}, 
we plot the bit error rate (BER) performance of NBLDPC coded $16 \times 16$ MIMO system with QPSK modulation $(M=4)$.
The turbo coded STBC (space-time block code) MIMO systems with LAS (likelihood ascent search) detector \cite{coded_large_mimo3} are chosen for comparison purpose. 
For the $16 \times 16$ MIMO system with QPSK modulation, 
these turbo coded systems perform within about 4 dB 
from the capacity which is known as the best performance founded in the literature.
By using $R=1/3$ and $R=1/2$ NBLDPC codes with $n=3456$ bits, 
NBLDPC coded systems outperform turbo coded systems by approximately 0.5 dB.
We also show the BER performance of short length NBLDPC codes $n=864$.
As we expected, the use of shorter length $n=864$ codes result in degraded BER performance.
In term of nearness-to-capacity, 
the performance of NBLDPC coded systems is 3.5 dB away from MIMO capacity.
We mention that the MIMO transmission can be classified into two main categories : 
1) \textit{spatial multiplexing} for higher data rate and
2) \textit{space-time coding} for higher transmission quality.
We, however, compare the turbo coded system \cite{coded_large_mimo3} which is space-time coding 
and the proposed NBLDPC coded system which is spatial multiplexing
since both have the same spectral efficiency (also MIMO capacity).

\begin{figure}[htb]
\centering
\includegraphics[scale=0.6]{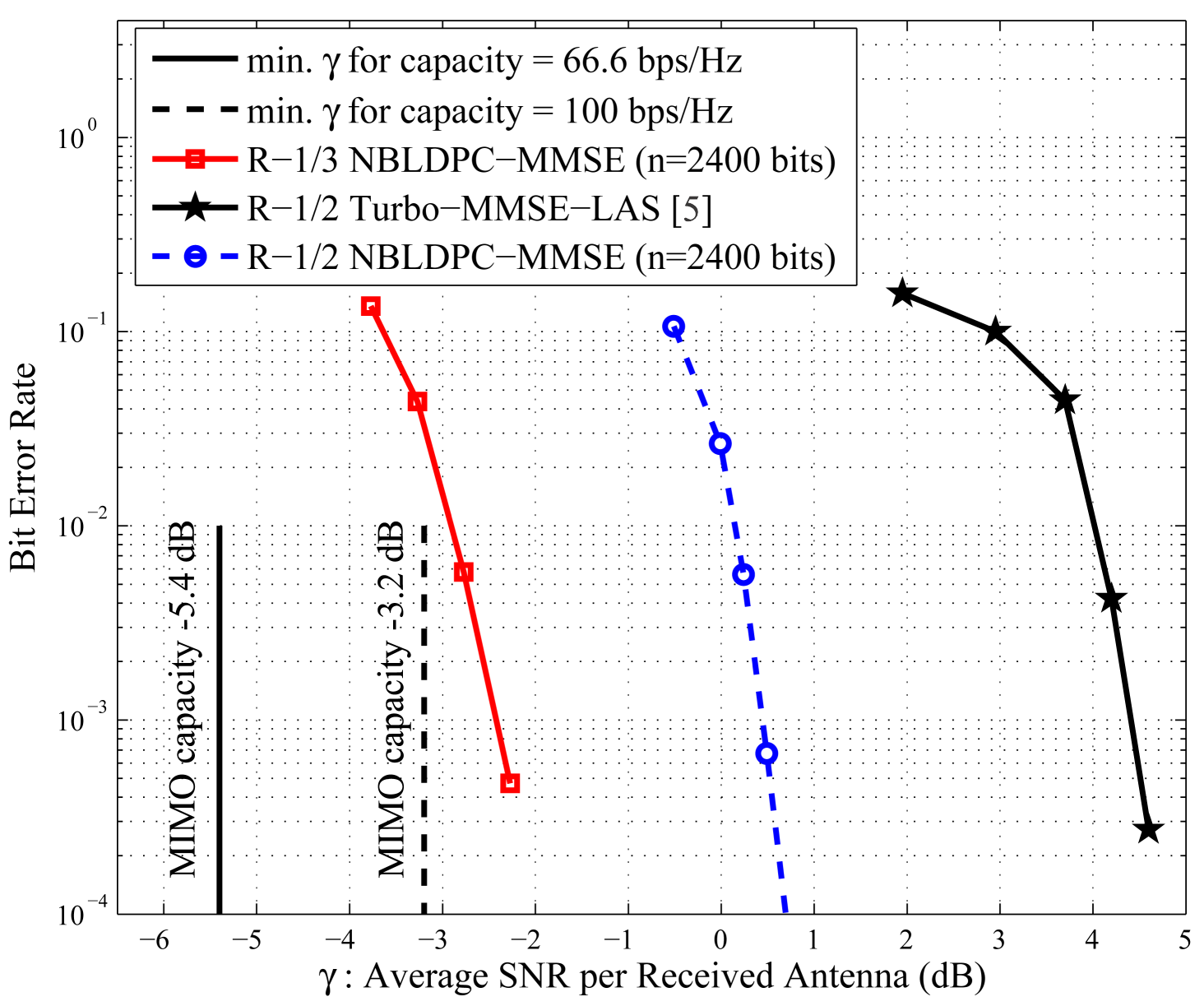}
\caption{Bit error rate curve of coded $200 \times 200$ MIMO systems with BPSK modulation.
The spectral efficiencies of 66.6 and 100 bps/Hz 
are obtained from MIMO system with channel codes of $R=1/3$ and $R=1/2$ respectively.}
\label{ber200x200}
\end{figure}

The BER performance of NBLDPC coded $200 \times 200$ MIMO systems 
with BPSK modulation $(M=2)$ is shown in Fig. \ref{ber200x200}.
For $200 \times 200$ MIMO system,
the best performing scheme which can be founded in the literature is
the $R=1/2$ turbo coded spatial multiplexing MIMO systems with the MMSE-LAS detector \cite{coded_large_mimo1}.
The MMSE-LAS detection algorithm uses the MMSE detection to initialize the algorithm.
Therefore, the overall computational complexity of MMSE-LAS detector is greater than that of MMSE detector.
The BER performance of this turbo coded system is away from MIMO capacity by 7.5 dB.
It is clearly seen from the figure that
$R=1/2$ NBLDPC coded system with $n=2400$ bits significantly outperforms turbo coded system by about 4 dB.
Both $R=1/3$ and $R=1/2$ NBLDPC coded systems perform within just 3.5 dB from the MIMO capacity

\begin{figure}[htb]
\centering
\includegraphics[scale=0.59]{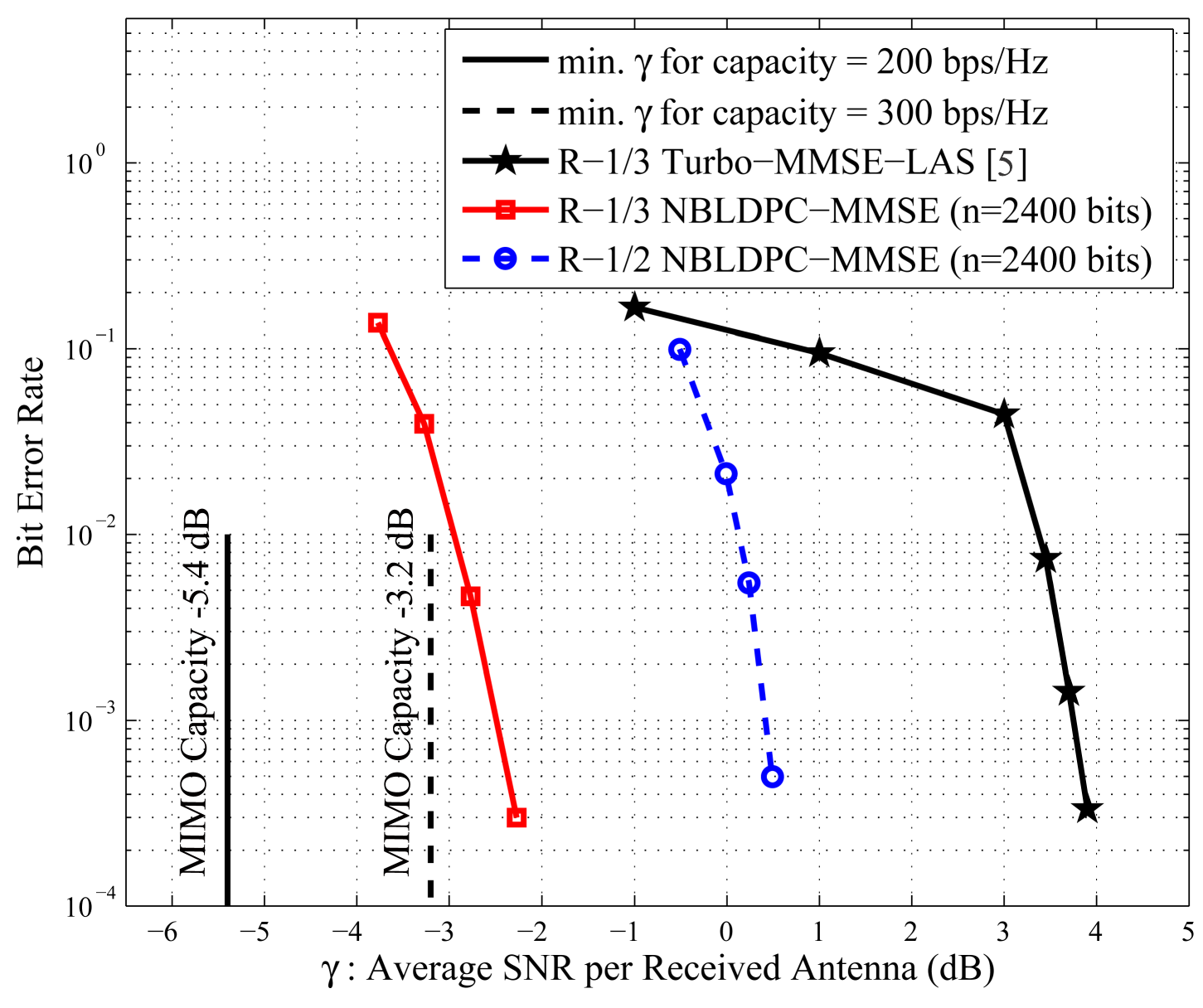}
\caption{Bit error rate curve of coded $600 \times 600$ MIMO systems with BPSK modulation.
The spectral efficiencies of 200 and 300 bps/Hz 
are obtained from MIMO system with channel codes of $R=1/3$ and $R=1/2$ respectively.}
\label{ber600x600}
\end{figure}

Figure \ref{ber600x600} presents the simulation results for larger dimension, $600 \times 600$ MIMO system.
It can be observed that $R=1/3$ and $R=1/2$ NBLDPC coded systems operate within
3.4 and 3.6 dB, respectively, from the corresponding MIMO capacities.
The turbo coded system with MMSE-LAS detector in \cite{coded_large_mimo1} is again used as the competitor.
The performance of this turbo coded system over $600 \times 600$ MIMO system 
is away from the capacity by about 9.4 dB.
It is obviously seen that $R=1/3$ turbo coded system underperforms $R=1/3$ NBLDPC coded system by more than 6 dB.
More interestingly, $R=1/2$ NBLDPC coded system absolutely outperforms $R=1/3$ turbo coded system by about 3.4 dB.
Therefore, the NBLDPC coded system outperforms the turbo coded system both in term of performance and spectral efficiency.

\begin{figure}[htb]
\centering
\includegraphics[scale=0.57]{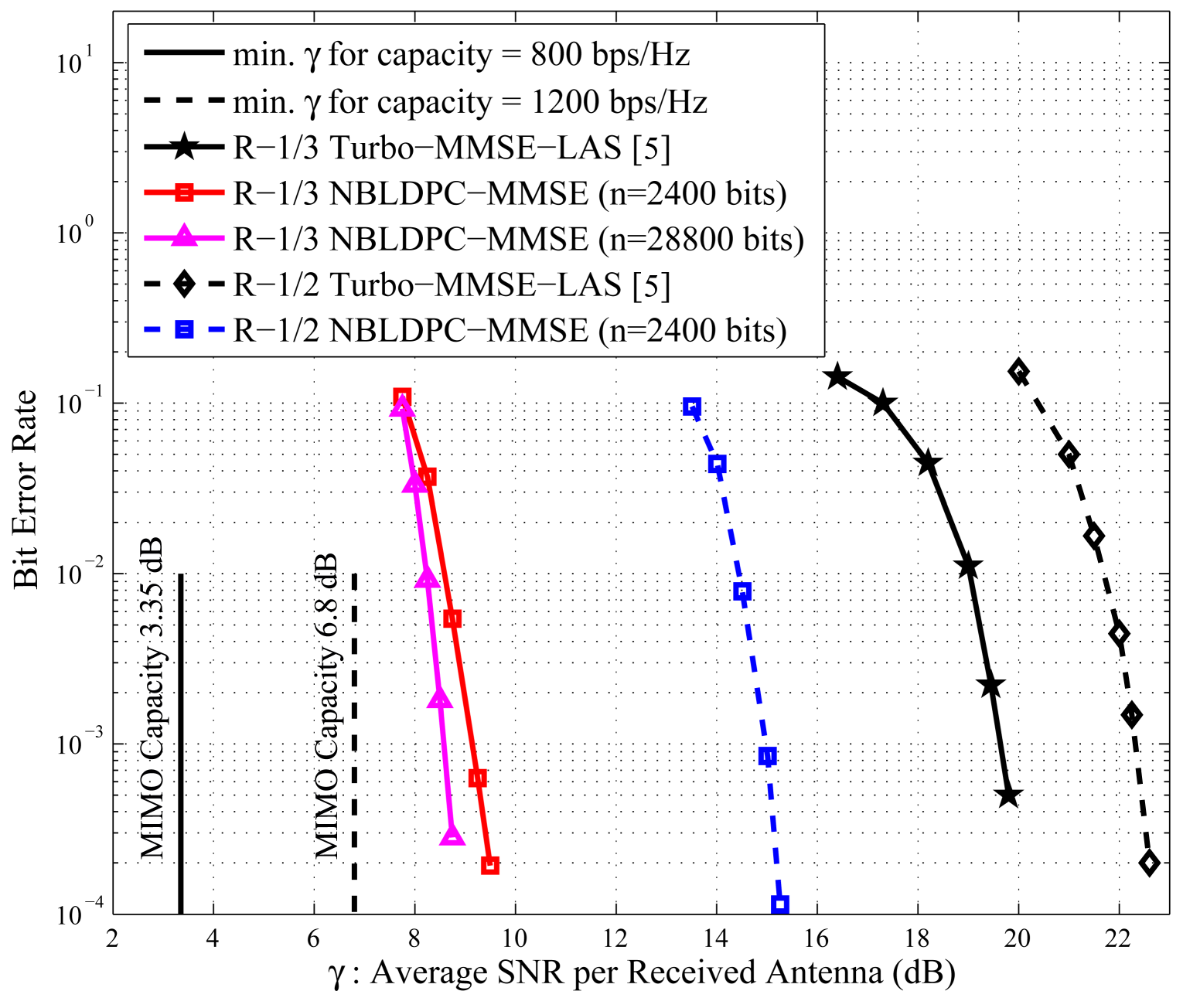}
\caption{Bit error rate performances of coded $600 \times 600$ MIMO systems with 16-QAM modulation.
The spectral efficiencies of 800 and 1200 bps/Hz 
are obtained from MIMO system with channel codes of $R=1/3$ and $R=1/2$ respectively.}
\label{ber600x600_16QAM}
\end{figure}

\begin{figure}[htb]
\centering
\includegraphics[scale=0.54]{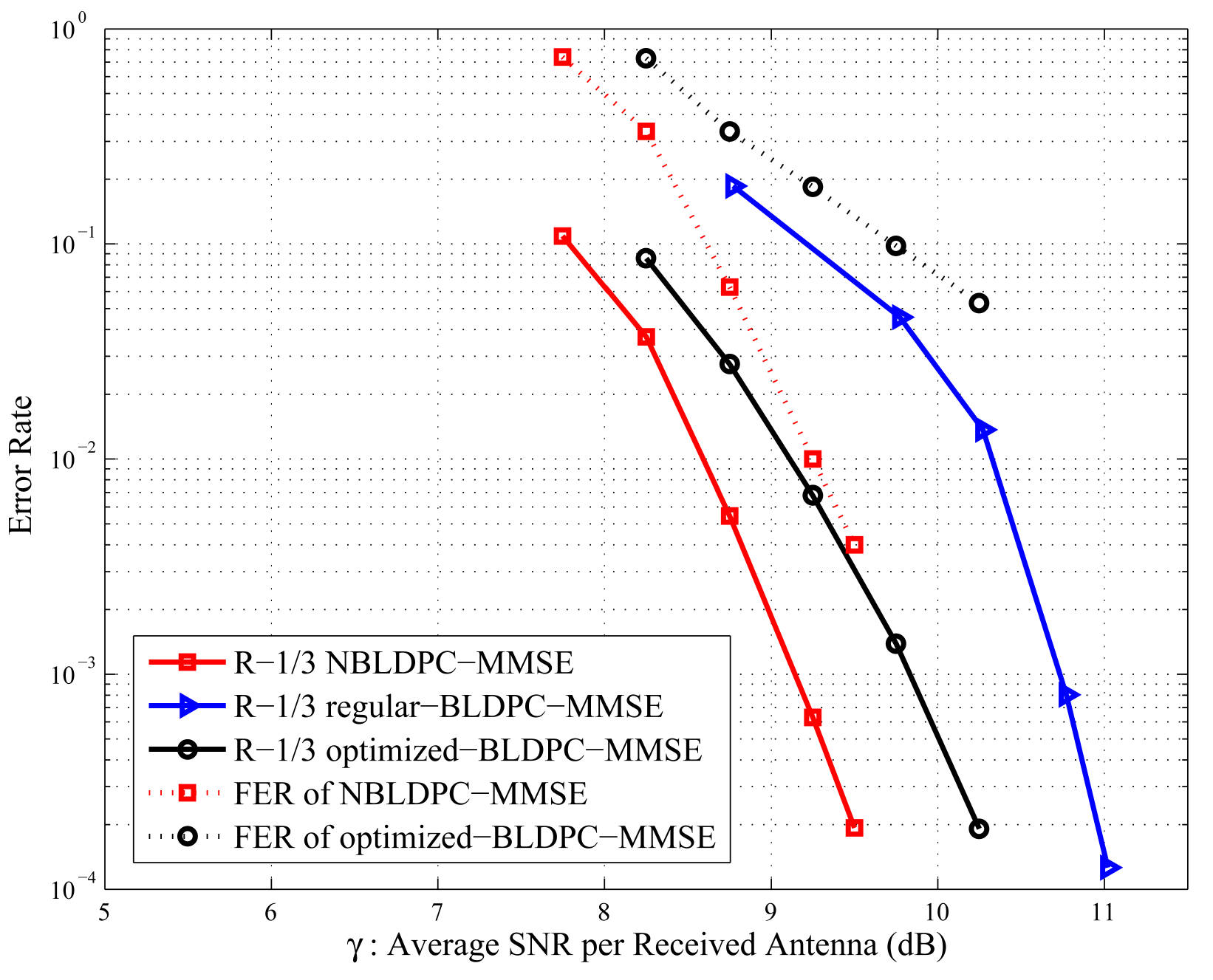}
\caption{Performance comparison between $R=1/3$ NBLPDC code and $R=1/3$ BLDPC codes on $600 \times 600$ MIMO systems with 16-QAM modulation at $n=2400$ bits.
The solid curves represent the BER performance while the dashed curves are the corresponding FER performance.}
\label{comparison600x600_16QAM}
\end{figure}

We also investigate the BER performance of coded large MIMO system with higher-order modulation.
The coded performance of $600 \times 600$ MIMO system with $16$-QAM is 
shown in Fig. \ref{ber600x600_16QAM} and \ref{comparison600x600_16QAM}.
From both figures, the following observations can be listed as follows :

\noindent$\bullet$ $R=1/3$ and $R=1/2$ NBLDPC coded systems with $n=2400$ bits
operate within 6 dB and 8 dB, respectively, from MIMO capacity.

\noindent$\bullet$ $R=1/3$ and $R=1/2$ turbo coded systems \cite{coded_large_mimo1} 
operate very far (more than 15 dB) from MIMO capacity. 
It is clearly seen that the NBLDPC coded systems indeed beat turbo coded systems. 

\noindent$\bullet$ By increasing the length of code from $n = 2400$ bits to $n=28800$ bits, 
the coding gain about 0.7 dB can be obtained from $R=1/3$ NBLDPC coded system.

\noindent$\bullet$ For moderate codeword lengths ($n=2400$ bits) and the same MMSE detector, 
$R=1/3$ NBLDPC coded systems outperform both the optimized and regular binary LDPC (BLDPC) codes. 
The coding gains obtained over optimized and regular binary LDPC codes are about 0.8 and 2 dB respectively.
In Fig. \ref{comparison600x600_16QAM}, the regular binary LDPC code has column weight 4 and row weight 6
whereas the optimized binary LDPC code is taken from Table. III in \cite{opt_bldpc}.
The maximum degree of variable node of the selected optimized binary LDPC code is 16.
This optimized LDPC code can asymptotically performs very close to 
the capacity of single input single output Rayleigh fading channel (within 0.19 dB).

\noindent$\bullet$ Another advantage of using NBLDPC code which can be seen from this figure
is the excellent frame error rate (FER).
Comparing with optimized BLDPC code, sharp decrease in the FER curve is obtained. 
Although the BER curve of optimized BLDPC code is good but its corresponding FER curve is quite bad.

For binary LDPC coded MIMO system with MMSE detector,
the soft output from detector is slightly different from NBLDPC coded system. 
We note that the input of binary LDPC decoder is the probability of size two 
which can be represented by log likelihood ratio (LLR).
The LLR for $l$th bit $\left(\forall l \in \lbrace 1,\ldots,p \rbrace\right)$ 
of the symbol $s_k$ transmitted from $k$th antenna is given as
\begin{align*}
 \mathrm{ln} \frac{\sum_{s \in  \mathbb{A}^{1}_{l} } \mathrm{Pr}\left( \hat{s}_k \mid s \right) }{\sum_{s \in  \mathbb{A}^{0}_{l} } \mathrm{Pr}\left( \hat{s}_k \mid s \right) },
\end{align*}
where $\mathbb{A}^{1}_{l}$ is the set of modulated symbols whose binary labelling of $l$th bit is 1 ($\mathbb{A}^{0}_{l}$ is similarly interpreted) and $\mathrm{Pr}\left( \hat{s}_k \mid s \right)$ is computed from (\ref{Prob_eq}).
It is known that the performance of BLDPC coded system can be enhanced 
by adding the number of iterations between decoder and detector 
but, for large MIMO systems, this will greatly increase the computational complexity.
Thus, the simulation for BLDPC coded system with joint detection and decoding is not carried out.

\begin{figure}[htb]
\centering
\includegraphics[scale=0.675]{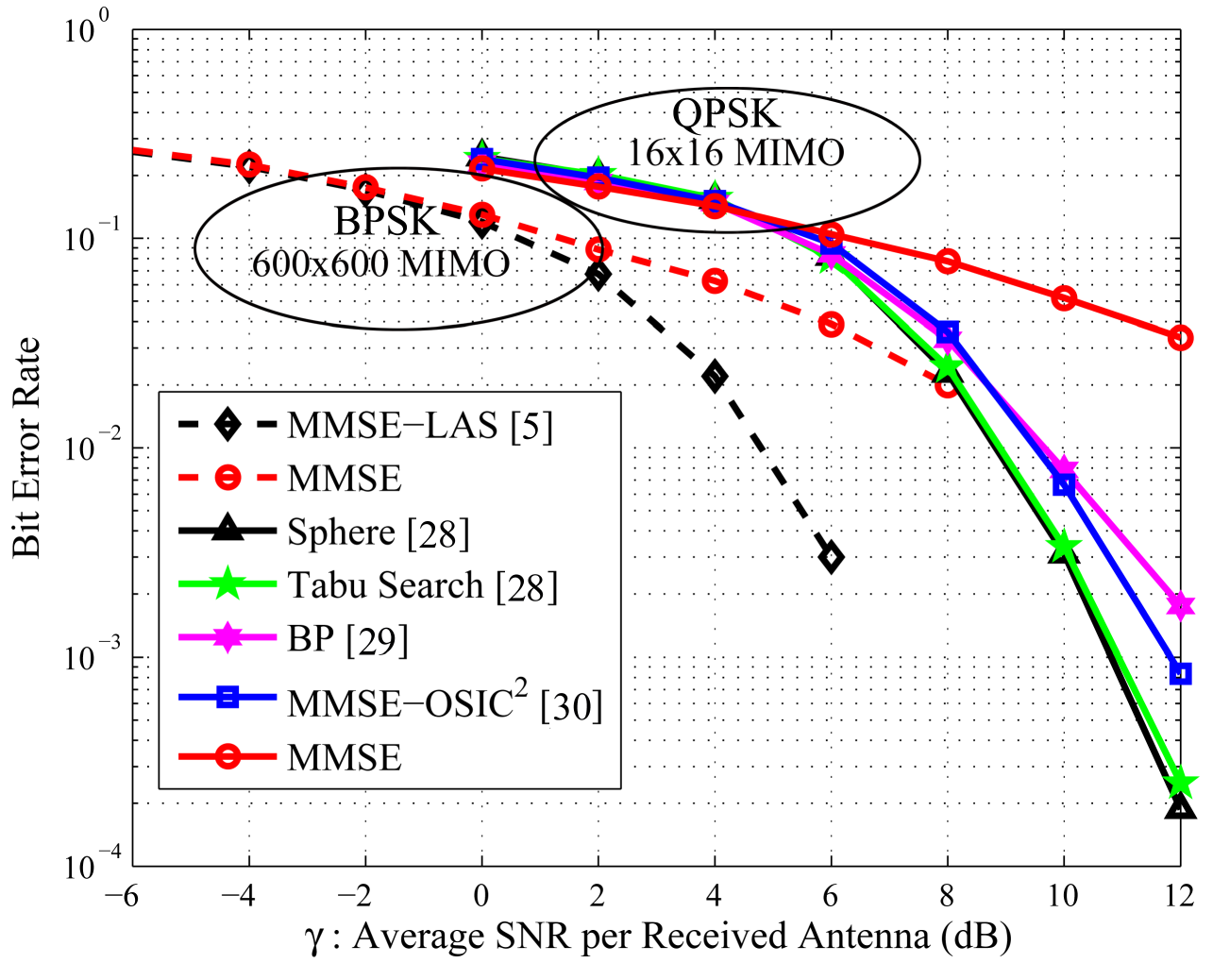}
\caption{Uncoded performances for $600 \times 600$ and $16 \times 16$ MIMO systems.
The results of MMSE-LAS detection \cite{coded_large_mimo1}, Tabu search \cite{tabu}, sphere detection \cite{tabu},
Belief propagation (BP) based detection \cite{BP}, MMSE-OSIC$^2$ detection \cite{mmse_osic2} 
which have very good performance in large MIMO systems
are included for comparative purpose.
The operating regions for coded system are highlighted by ellipses.}
\label{uncoded_BER}
\end{figure}

Although the MMSE detector has very poor detecting performance for MIMO system with $N_t = N_r$
but in this section we have demonstrated that 
the application of this detector to NBLDPC coded system is somewhat excellent. 
The uncoded performance of MMSE detector is extremely poor comparing to the optimal one. 
The MMSE detector has a diversity order of $N_{r} - N_{t} + 1$ 
while the optimal ML detector has a diversity order of $N_r$ \cite{coded_large_mimo1}.
The MMSE detector thus achieve only first order diversity for MIMO system with $N_t = N_r$.
Therefore, One may be surprised why MMSE detector yields such good coded performance.
Let us take a closer look at the uncoded results  shown in Fig. \ref{uncoded_BER}.
It can be observed that the uncoded performance of MMSE detector 
for $600 \times 600$ and $16 \times 16$ MIMO systems
are very similar to those of near-optimal MIMO detectors for low SNR region.
Noticeably, those regions are the operating regions for coded systems.
By trial and error, 
we suspect that the key component to achieve good performance from NBLDPC coded system
is not a detection part but a soft-output generation.
We note that other methodologies for generating soft-output from linear MIMO detection
such as the one described in \cite[p.~358]{soft_out} 
leads to seriously degraded performance (not shown here).

Based on the simulations provided in this section,
we list our major results as follows :

1) The BER performances obtained by NBLDPC coded large MIMO systems are now the best known performance
and very useful for further investigations.

2) In term of BER performance, the NBLDPC coded systems outperform turbo coded systems and also binary LDPC coded systems.
The provided results reveal that the turbo code may not be a powerful channel code for large MIMO systems
especially when the higher order modulation is adopted.

3) The application of MMSE detector to coded large MIMO systems is very promising
if the operating region is near the MIMO capacity.

\section{Conclusion}
In this paper, the NBLDPC coded large MIMO systems are studied.
The low-complexity MMSE detector is employed to provide the soft-input for NBLDPC decoder.
We have demonstrated that the proposed NBLDPC coded system 
can significantly decrease the remaining gap from MIMO capacity 
which is previously obtained from the best known turbo coded systems.
By using moderate length NBLDPC codes (a few thousand bits), 
the proposed coded system can perform near MIMO capacity which is closest than ever.
We therefore conclude that the NBLDPC coded large MIMO system can be one of the best choices
to achieve both the excellent BER performance and the ultra high spectral efficiency.


\section*{Acknowledgments}
This work is financially supported by the Telecommunications Research Industrial and Development Institute (TRIDI), with National Telecommunications Commission (NTC), Grant No.PHD/009/2552. 
The authors also would like to acknowledge the discussions and guidance of Rong Hui Peng.

\bibliographystyle{IEEEtran}
\bibliography{IEEEabrv,my_references}

\end{document}